\documentclass[%
 reprint,
 amsmath,amssymb,
 aps,
]{revtex4-2}
\usepackage{comment}
\usepackage{svg}
\usepackage{graphicx}
\usepackage{dcolumn}
\usepackage{bm}

\newcommand{\e}{\mathrm{e}}
\newcommand{\C}{\mathrm{C}}
\newcommand{\Hi}{\mathrm{H}}

\begin{document}

\preprint{}

\title{Thermodynamics of prebiotic synthesis}
\author{Michael Grayson}
 \altaffiliation{
 National Institute of Standards and Technology, Boulder, CO 80305,USA}
\date{\today}

\begin{abstract}
In this paper, I outline the correspondence of vector partitions with the combinatorics of dynamical systems whose states are partitions of multisets. I use the notion of a limiting shape of vector partitions to determine properties of the equilibrium distribution for these systems. I investigate prebiotic systems as ergodic series of chemical reactions as an example. I determine under what conditions long polymers of RNA are favorable at equilibrium using the limit shape of the vector partitions.
\end{abstract}

\maketitle


\section{Introduction }

Synthesis refers to the process of combining simpler substances or elements to form a more complex product. This includes processes such as nucleosynthesis of the elements, prebiotic synthesis, phonetic synthesis, and many more. Synthesis is the natural process by which complexity emerges in our universe. The mathematical models used in this paper describe the combinatorics and thermodynamics of synthesis—the aggregation of simple elements into complex products. This aggregation is driven by changes in indistinguishability. The loss of indistinguishability during aggregation increases entropy, which drives the synthetic process. In this paper, we will specifically apply this model to prebiotic synthesis. However, this theory can be applied to any situation where indistinguishable elements combine to form complex ones in an ergodic way. The results from this paper are a preliminary outline of what can be deduced from current research into integer partitions and their limit shapes. This paper represents the first step towards this goal and can be summarized into this simple statement:

\begin{center}
\textit{Entropy drives simple indistinguishable elements to form complex aggregates.}
\end{center}

This represents a philosophical deviation from the norm. It is common-sense intuition that diffuse systems of particles represent high entropy states \cite{schroeder_introduction_2000}. However, these states only consider the multiplicity of configurations of unbound particle in space. Over long enough time scales, all particles interact; even atoms fuse and decay into differently sized elements. Scaling time can be viewed as an effective increase in temperature, allowing the system to surpass higher activation energy barriers over longer timescales\cite{milburn_thermodynamics_2020}. This is due to the fact that improbable reactions become more likely over extended periods, as described by Poisson statistics, which can be interpreted as analogous to a higher temperature in the Arrhenius equation. When we actually look for the maximum entropy in the ergodic case, we find a tendency for an increase in complexity. That complexity is not an accident but a thermodynamic imperative. An entropic aggregation force arises much like the case of elastomers or self organizing lattices \cite{frenkel_entropy-driven_1999, ortiz_entropic_1999,van_anders_understanding_2014}. I do not claim that this is the fundamental drive for complexity; for example, natural selection, the tendency of systems that copy themselves well to exist for longer times, is a completely different process \cite{vanchurin_thermodynamics_2022}. Neither is this theory applicable to the details of how these complex systems behave. This theory applies to systems of aggregation, with such chaotic and varied dynamics, that they represent a random walk through configuration space. It provides tools for making long-time scale, ensemble predictions \cite{moore_ergodic_2015}. To provide motivation for this concept we will look at the origins of life.

It is a commonly held belief that life exists as a result of improbable events. However, there is a growing consensus among the academic community that this is not the case \cite{robertson_origins_2012, higgs_rna_2015}. Abiogenesis refers to the natural process of life arising from non-living matter, such as simple organic compounds. Can we find a physical theoretical justification for this process? Under what conditions does it exist, and what is its strength? Abiogenesis can be broken down into two steps: a cascade of random reactions that form larger compounds from smaller elements or molecules, and the subsequent proliferation of self-replicating molecules. The first process is known as prebiotic synthesis, while the second is self-replication. Both of these principles have been demonstrated experimentally. The famous Miller-Urey experiment demonstrated that complex organic molecules spontaneously form under simulated conditions believed to resemble the early Earth \cite{mccollom_miller-urey_2013,ferus_formation_2017}. Sidney Altman and Thomas Cech demonstrated in their work the ability of RNA to catalyze reactions and self-replicate \cite{miele_autocatalytic_1983, horning_amplification_2016, higgs_effect_2016}. While we now have some understanding about how abiogenesis may have happened, we are still lacking a complete physical model. For example, it is still unclear how long strands of RNA could spontaneously form, as it is not stable in water \cite{nam_abiotic_2018, cafferty_abiotic_2014}. 

The vast majority of research in this field concerns itself with determining the particular pathways by which abiogenesis can occur \cite{higgs_rna_2015, robertson_origins_2012}. I will be taking a far more coarsely grained approach. I will look at, in the most general case, what drives synthesis to occur in general and specifically apply it to prebiotic synthesis. Recently, it has been demonstrated that the distribution of types of molecules in prebiotic conditions, such as the Miller-Urey experiment or the Murchison meteorite, generalizes to the distribution of all organic molecules that have been recorded \cite{kauffman_theory_2020, wollrab_millerurey_2018}. This distribution is governed by the combinatorics of integer vectors whose indices correspond to atomic masses present and values correspond to the number of masses or atoms \cite{kauffman_theory_2020, wollrab_millerurey_2018}. These studies are limited in that they only consider the number of different compounds, but they do not consider the multiplicity of these different molecules \cite{kauffman_theory_2020, wollrab_millerurey_2018}. I will include this multiplicity and demonstrate in the ergodic case that there is a thermodynamic incentive to produce long strands, given a sufficient concentration of precursors.

In particular, we will be looking at the combinatorics of turning simple compounds into complex molecules. We will consider the case where particles are just as likely to break up as they are to aggregate so that we can apply our combinatoric model. The general time-dependent case is governed by the Smoluchowski coagulation-fragmentation equation, which is an infinite set of nonlinear differential equations \cite{chen_why_2022, jeon_existence_1998}. We will get around the task of solving such a set by assuming the system is ergodic. We can then simply look at the distribution of states in the phase space or their entropy.

Usually, it is assumed that the aggregation proceeds irreversibly, which gives a steady-state solution known as the Flory-Schulz distribution \cite{spaeth_polyaddition_2020}. This distribution is essentially a Poisson distribution over aggregate sizes. This distribution models well the dynamics of polymerization over short time scales when polymers do not break down. For example, RNA synthesis reactions that occur rapidly over the course of days \cite{higgs_effect_2016, nam_abiotic_2018}. However, over long time scales, any type of reaction that can happen will happen. RNA will break down in water due to hydrolysis, and under the best-case scenario, is stable for a few months \cite{fabre_efficient_2014}. The evolution of prebiotic compounds is concerned with hundreds of millions of years \cite{higgs_rna_2015}. Additionally, at these time scales, the climate of the Earth is a chaotic system that will expose these molecules to extreme perturbations across every degree of freedom: temperature, pressure, pH, solvent selection, etc. The combination of these effects will drive our system to occupy every state in its phase space. One of the difficulties of understanding prebiotic synthesis is the varied environments these molecules face, from extreme cold to meteor impacts \cite{ferus_formation_2017, yadav_chemistry_2020}. In our model, we take advantage of this variety by invoking ergodicity, assuming their effects are strong, varied, and chaotic. This is equivalent to the assumption that over long time periods, the system will be at equilibrium. In addition to ergodicity, we will need to assume molecules are to some extent indistinguishable.

It is well known that molecules of the same type, particularly simple ones, behave as if they are indistinguishable \cite{schroeder_introduction_2000}. This somewhat surprising fact means that the states of the system are the same under the exchange of any two particles. This is known as exchange symmetry. Materials that do not have exchange symmetry will have non-extensive material properties, particularly entropy, i.e., not additive \cite{schroeder_introduction_2000}. This is due to an additional $N!$ amount of states that must be counted. For example, doubling the volume of an ideal gas of distinguishable particles would increase its entropy by $2\ln(2)$. This paradox, that particles must be indistinguishable, was discovered in 1875 by Josiah Gibbs. In the modern viewpoint, it is not so much that the particles cannot be distinguished through any means; rather, it is that the dynamics of the system do not distinguish the particles. It is therefore a perfectly valid assumption that these simple interacting compounds be indistinguishable.

\section{Mathematical Framework}
\begin{figure}
\includesvg[width=0.5\textwidth]{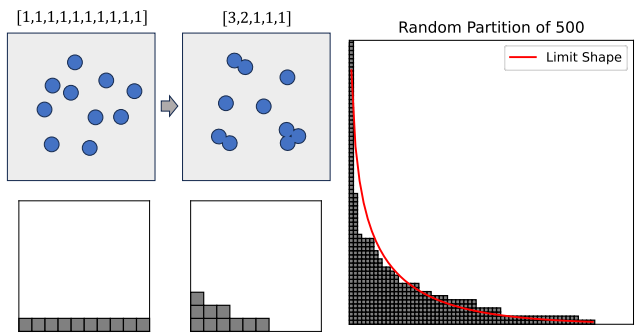}
\caption{\label{fig:limit} An image showing the atomic partition of 10, representing 10 unbound simple compounds. A transition into a random partition of 10 is very likely if final states are chosen at random. Bottom shows the Young diagrams for these partitions. On the right is a random partition of 500 showing that it approaches the limit shape. }
\end{figure}

The probability of an event occurring is determined by its change in entropy. This principle, known as the fluctuation theorem, essentially restates the second law of thermodynamics \cite{evans_fluctuation_2002}.
\begin{equation}
\frac{P(A \rightarrow B)}{P(B \rightarrow A)} = \e^{\Delta S_{AB}/k_{B}}
\label{eq:entropy_relation}
\end{equation}
The probability of transition from A to B is exponentially more likely than its reverse as entropy increases \cite{evans_fluctuation_2002}.
To calculate the entropy of a state, we need to count the multiplicity of the macro-states \cite{schroeder_introduction_2000}. In this model, we will have simple indistinguishable compounds that can bind to form new, more complex compounds which are distinguishable. We will start with only one type of indistinguishable compound. Consider a system of $4$ indistinguishable particles which can bind. We will label the indistinguishable particles as $1$ and group two particles by $2$, and so on. Let us list all possible configurations; for now, we will ignore the order and the empty spaces the particles may occupy.
\begin{align}
\{1,1,1,1\},\:
\{2,1,1\},\:
\{2,2\},\:
\{3,1\},\:
\{4\}
\label{eq:set_partition}
\end{align}
Figure \ref{fig:limit} illustrates this process for $n$ particles aggregating from the atomic state to a random partition \cite{fatkullin_limit_2018}. These represent all the possible ways to add integers to obtain $4$, also known as the integer partitions of $4$. Figure \ref{fig:limit} depicts two partitions of $10$ and how they can be arranged into a Young diagram. The field of integer partitions is a rich and well-studied area of mathematics, intimately related to the statistics of particles \cite{vershik_fluctuation_2006, maeda_amoebas_2007, cerf_low-temperature_2001}. Integer partitions have a special property: a random partition of a large integer generally has the same distribution \cite{fatkullin_limit_2018}. This property is known as the limit shape. The limiting shape of the integer partitions, $F(x)$, was originally determined by Vershik \cite{vershik_statistical_1996}.
\begin{equation}
\e^{\frac{-\pi x}{\sqrt{6}}} + \e^{\frac{-\pi y}{\sqrt{6}}} = 1
\label{eq:lim_shape}
\end{equation}

\begin{equation}
\rho(x) = -F'(x) =  \frac{1}{\e^{\frac{\pi x}{\sqrt{6}}} - 1}
\label{eq:lim_density}
\end{equation}

Where Eq. \ref{eq:lim_shape} is the limiting shape of a normalized young diagram, and Eq. \ref{eq:lim_density} is its corresponding probability distribution. The derivation and definitions are in Appendix \ref{app:1}. Unsurprisingly, the statistics of indistinguishable elements follow boson statistics \cite{vershik_statistical_1996, fatkullin_limit_2018}. Taking a power series expansion around zero, we find

\begin{equation}
p(x) \approx \frac{\sqrt{6}}{\pi} \frac{1}{x} - \frac{1}{2} + \frac{\pi x}{12 \sqrt{6}} + \dots
\label{eq:approximation}
\end{equation}

Which is dominated near $0$ by the powers series $x^{-1}$. This power law, commonly known as Zipf's Law or the Zipf-Mandelbrot Law, is a universal feature in many systems that consist of partitioning some amount of resource into subgroups. This includes the distribution of word frequency in texts, gene frequency in genomes, species abundance in an ecosystem, marine particle size distribution, etc. \cite{bender_genetic_1986, kalankesh_language_2012, nicolis_chaotic_1989, piantadosi_zipfs_2014, mcgill_species_2007}. Zipf's Law is a very universal distribution, and it is unlikely that a single explanation is responsible for its existence. It is more likely that it is part of a universality class with many underlying models.

The integer partitions have the property that for large $n$ nearly all the partitions of $n$ are "close" to the limit shape. Figure \ref{fig:limit} depicts the limit shape in red along with a randomly chosen partition of 500 in black. Therefore the multiplicity of partitions "close" to the limit shape will be on the order of the total number of partitions of the interger $n$. The number of partitions for an integer $n$ is given by the Hardy-Ramanujan formula \cite{fatkullin_limit_2018}. 
\begin{equation}
p(n) \approx \frac{1}{4n\sqrt{3}}\exp\left(\pi\sqrt{\frac{2n}{3}}\right)
\label{eq:partition_function}
\end{equation}

We can split up our system into two macro-states: partitions that are "close" to the limit shape, and those that are not. A simpler choice would be to use the ordering of the partitions, known as the integer compositions \cite{yakubovich_coincidence_2005}. However, this would result in a non-extensive entropy \cite{jensen_statistical_2018}, see appendix 
\ref{app:2}. The entropy of the limit shape will be $S \approx k_{B}\ln(p(n))$, the ln of the number of partitions.  The entropy of the other macro-state is just $k_{B}\ln(1) \approx 0$. For a more rigorous definition of macro-state we should use the entropy defined in the variational derivation in appendix \ref{app:1}. Therefore the increase in entropy of our system is $k_{B}\ln(p(n))$ which is also represents the system's maximum entropy. The probability of transitioning from some particular state $A$ to the limit shape $B$ is then
\begin{equation}
P(A \rightarrow B) = \e^{k_{B}\ln(p(n))/k_{B}}P(B \rightarrow A)
\label{eq:trans_prob1}
\end{equation}

The probability of transitioning from \(B\) to the limit shape is just \(1/p(n)\) because it is one state out of \(p(n)\) states.

\begin{equation}
P(A \rightarrow B) = \frac{\e^{\ln(p(n))}}{p(n)}
\label{eq:trans_prob2}
\end{equation}

\begin{equation}
P(A \rightarrow B) \approx 1
\label{eq:trans_prob3}
\end{equation}

Which of course converges to $1$. This is of course a gross oversimplification, but in the limit of large $n$ these simplifications are true. The probability of the system evolving to the limit shape is unity. The rate at which this probability converges to $1$ is on the order of $\ln(p(n))$ or $\sqrt{n} $. It is certainly possible to derive a probability for a particular $n$, but this is not a simple task. A measure of closeness is defined in appendix \ref{app:1}. For now we will limit ourselves to approximating it as unity. 

This means that indistinguishability drives aggregation. The aggregation in this model is completely fictitious; it is an emergent entropic force. This entropic force is in the same universality class as the restoring force of an elastomer (rubber band)\cite{ortiz_entropic_1999, nahum_quantum_2017}. The origin of this force does not come from the particles themselves but from how the environment they exist in coarsely grains them. This entropic force will be able to overcome energy barriers on the order of $kT\ln(p(n))$. There always exists a temperature at which this force will dominate. If we were to specify that every particle was distinguishable, we would see no aggregation. This situation is equivalent to Bell statistics, whose limit shape is the indicator function, i.e., no particles are bound together \cite{fatkullin_limit_2018}. Additionally, the limit shape will be affected if we require that molecules are built in an orderly manner; this is the Plancherel measure \cite{fatkullin_limit_2018}.

We have, therefore, that the tendency of particles to form aggregates is driven by entropy. This process converts simple indistinguishable particles into complex distinguishable ones, thereby producing entropy. The fictitious force of aggregation that particles will experience is on the order of $k_{B}T\ln(p(n))$. The conditions that favor the formation of complex molecules from smaller ones are hot baths of many indistinguishable simple compounds.

We can additionally provide bounds on the size of the aggregates in the partitions. The largest particle in the limit partition is $\sqrt{n}\ln{n}$ and and there are very few particles of this size.\cite{vershik_fluctuation_2006, erdos_distribution_1941}. Since the limit shape is symmetric, this is also the number of of total molecules. So, we can not only roughly predict the conditions under which more complex molecules will form, but we can also predict their maximum size. This concludes what I believe to be the most important points of partition theory in the context of abiogenesis. Of course, there are many more principles of the partitions to be applied, but this is enough for us. To move further, we must move on to a different type of partition, one that is slightly more complex.

Usually, we do not have a single indistinguishable compound. For example, RNA is formed of the 4 nucleotides, which are distinguishable between each other \cite{higgs_rna_2015, robertson_origins_2012}. Consider the situation of hydrocarbons. There are two types of indistinguishable particles: $\text{C}$ and $\text{H}$. For now, we will ignore the rules of chemistry. Let us look at the possible ways they can bond. Consider the molecules that could form from $\text{2H}$ and $\text{2C}$. We have
\begin{align}
&(2\C,2\Hi),\:(2\C,\Hi_{2}),\:(\C,\Hi,\C\Hi),\:(\Hi,\C_{2}\Hi),\:(2\C,\Hi_{2}) \nonumber \\
&(\C_{2},\Hi_{2}),\:(\C,\C\Hi_{2}),\:(\C\Hi,\C\Hi),\:(\C_{2}\Hi_{2})
\label{eq:chemical_species}
\end{align}

I have suggestively labeled these such that some nearly represent real molecules. This pattern has an exact correspondence to the partition of a integer vector. We can express this space of possibilities as the partitions of the multi-set $\{\C,\C,\Hi,\Hi\}$ \cite{bogachev_limit_2014,vershik_statistical_1996}. The partitions of the multi-set are represented by the vector partitions. These are the number of ways to express a vector $v$ as a sum of integer vectors $v_{k}$. For example the vector partitions of $[2,2]$ are
\begin{align}
&([0, 1], [0, 1], [1, 0], [1, 0]),\; ([0, 1], [0, 1], [2, 0]) \nonumber \\
&([0, 1], [1, 0], [1, 1]),\; ([0, 1], [2, 1]) \nonumber \\
&([0, 2], [1, 0], [1, 0]),\; ([0, 2], [2, 0]) \nonumber \\
&([1, 0], [1, 2]),\; ([1, 1], [1, 1]),\; ([2, 2])
\label{eq:tuple_groupings}
\end{align}

Calculations are performed using SageMath \cite{stein_sage_nodate}. If we image the first index counts the number of $\Hi$ and the second the number of $\C$ we can see the correspondence. for example  
\begin{equation}
    \:(2\C,\Hi_{2})\rightarrow  ([0, 1], [0, 1], [2, 0])
\end{equation}

The vector partitions are slightly more complex than the integer partitions making the derivation of their limit shape more difficult. As of yet there is not a complete derivation of the limit shape of the vector partitions. Luckily, its general properties are known. The number of vector partitions of a vector $[n,n,n,..,n]$ with dimension $d$ is approximately \cite{buffiere_asymptotic_2023, vershik_statistical_1996,barany_convex_2018}
\begin{equation}
p(n) \propto \exp\left(n^{\frac{d}{d+1}}\right)
\label{eq:partition_growth}
\end{equation}

Similar to the integer partitions case, we can estimate the largest vector (or molecule) in the limiting distribution to grow as

\begin{equation}
n_{\mathrm{max}} \propto n^{\frac{1}{d+1}}
\label{eq:largest_molecule}
\end{equation}

Both these equations are to leading order; the full equations have been derived for the number of zonotopes, which correspond to the strict vector partitions, but the leading order scaling in $n$ is valid \cite{buffiere_asymptotic_2023,barany_convex_2018}. The statistics of zonotopes are a natural geometric version of the vector partitions. They are convex polytopes whose boundaries vectors are described by the the strict partitioning of a vector. It should be noted that the rate at which these partitions grow is far more rapid than that of the integer partitions. In Appendix \ref{app:3}, I numerically estimate these growths a bit more accurately. Both of these equations are powerful summaries of the statistics of multiset partitions. For example, we could predict (roughly) the increase in entropy from the aggregation of a few simple, indistinguishable particles into larger, more complex ones. Additionally, we can predict what the largest possible molecule is that could be formed through aggregation. Many of the molecules listed in the example do not exist. There are two methods to elaborate this model. We could identify unstable combinations as metastable states and weight their likelihood using Boltzmann statistics \cite{fatkullin_limit_2018}. Alternatively, we could calculate the partitions in a completely different geometry, removing those that are impossible configurations \cite{maeda_amoebas_2007}. I think ideally a combination of both is required, as there are many metastable molecules that must form for reactions to take place. These molecules additionally make only a small transition to a stable state, resulting in a small perturbation to the limit shape.

As the dimension of the vector partition increases, distinguishability is added to our system. This indistinguishability in turn reduces the maximum degree of aggregation that can occur. In the limit of a dimension equal to the number of particles, the limit shape will approach that of the Bell measure, i.e., no aggregation at all. The vector partitions are limited to integer dimensions, and these results only apply for vectors where the number of elements is uniformly distributed over the dimension. However, we can map a nonuniform distribution of simple elements to a uniform one using entropy. Entropy, in information theory, represents the minimum number of uniformly likely symbols required to encode a message. If we have a distribution of abundances over different elements, we can calculate its entropy to see the minimum number of uniformly likely elements to use. First, we calculate the entropy over our elements $p_{i}$ and then we find a system of uniform elements $n_{\mathrm{e}}$ with equal entropy.
\begin{align}
H &= -\sum_{i}^{m} p_{i}\log_{2}(p_{i})
\label{eq:entropy_general} \\
H &= -n_{\mathrm{e}}(1/n_{e})\log_{2}(1/n_{\mathrm{e}})
\label{eq:entropy_specific} \\
n_{\mathrm{e}} &= 2^{H}
\label{eq:effective_states}
\end{align}

\begin{figure}
\includesvg[width=0.5\textwidth]{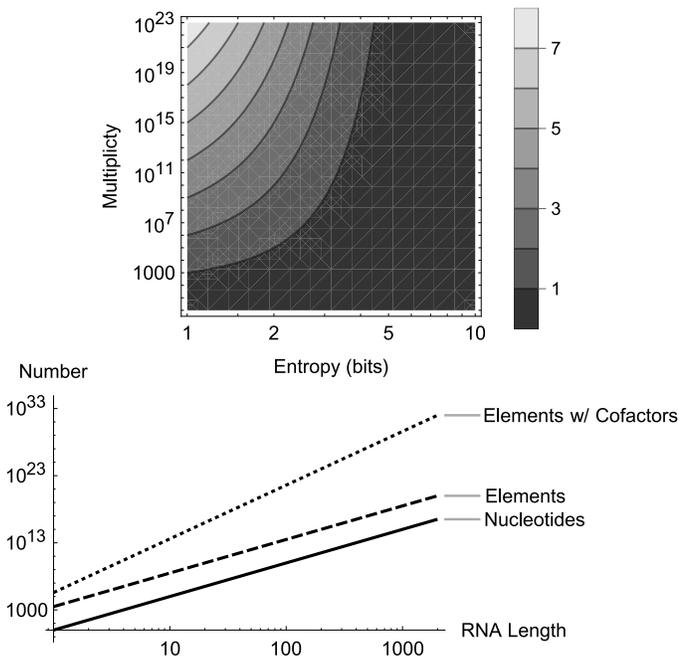}
\caption{\label{fig:vec} in the top is a depiction of the ln maximum element size as a function of multiplicity and dimension. This proportionality is just $n^{1/(d+1)}$. As systems become more distinguishable they tend to aggregate less. bottom, a depiction of the maximum length RNA molecule at equilibrium for various dimensions on a ln ln plot.  }
\end{figure}

Figure \ref{fig:vec} depicts the logarithm of the maximum size of a polymer as a function of Shannon entropy and multiplicity. As the distinguishability or entropy of the monomers increases, the incentive to bind into large polymers decreases. At approximately 5 bits, or 32 different characters, the maximum length polymer for a mole of monomers becomes less than 10. This plot is applicable to any system that fits the combinatorial problem of integer partitions. For example, integer partitions solve the problem of constructing a set of words that encode the most information, given a finite number of indistinguishable characters. If we include the ordering of the characters, known as integer compositions, the number of partitions increases greatly, but the limit shape is still the optimal encoding \cite{yakubovich_coincidence_2005}.

An important point is that the degree to which the partitioning model can be applied is determined by the coarse-graining applied by the systems they interact with. When partitioning models can be cascaded, it will result in larger aggregations than expected. These processes would be more accurately modeled by higher-dimensional partitions which include this cascading effect implicitly \cite{kenyon_limit_2007}.

\section{Conditions for pre-biotic synthesis}
Now that we have a simple model for aggregation in multi-component systems, we can apply it to a real-world scenario. In particular, we could ask, for example, what would be the required molarities of elements to possibly produce a piece of RNA that can self-catalyze? Currently, Miller-Urey experiments have been able to produce amino acids and nucleotides but have yet to produce self-replicating RNA molecules \cite{ferus_formation_2017, mccollom_miller-urey_2013}. This would be approximately 200 base pairs long \cite{johnston_rna-catalyzed_2001}. An important prediction of this theory is that the total size of the experiment determines the maximum sized molecules that will form. These volume dependencies on aggregate size are not uncommon in nanoparticle chemistry \cite{gao_facile_2016}. Therefore, the limitation of Miller-Urey experiments in producing complex molecules may only be a matter of scale. Using the theory explained in this paper, we can calculate the required number of precursor atoms and the entropy change produced by the aggregation process. These calculations are shown in Fig. \ref{fig:vec}. However, an additional calculation is the time and applied perturbations to the systems to guarantee the dynamics are ergodic. Additionally, these estimates only represent a lower bound for the required number of precursor atoms.

In the worst case scenario, this would require elements and cofactors for catalysis which roughly corresponds to a dimension of approximately 7 \cite{higgs_rna_2015, robertson_origins_2012}. Additionally, each nucleotide consists, on average, of 5 elements.  Which is about 2 mols of each element all interacting with each other. For a realistic experiment, the concentration of dissolved carbon in the ocean is $2.3\times 10^{-3}$ mol/kg \cite{zeebe_carbon_2009}. A realistic simulation of this system would then require approximately a metric ton of solution. Of course, the concentration is usually greatly enhanced in these experiments over that of sea water. The original experiment used about $.25$ mol/kg of methane to water with a total of $0.2$ kg water, more recent experiments used about $.2$ mol/kg This reduces the required concentration by 10, still a massive scale compared to current experiments \cite{parker_conducting_2014, mccollom_miller-urey_2013, ferus_formation_2017}. The free energy change associated with aggregation is only about 4 J at 300 Kelvin, minuscule amount compared to the energies involved in reactions. The activation energies of these reactions are on the order of tens of kJ/mol \cite{ferus_formation_2017}. For the purely entropic force approach 10 kJ/mol we would need approximately 100 mols of each precursor. This would have an equilibrium maximum length of about 400. It is important to note that the entropy in this system is not extensive, a physical phenomena that has yet to be experimentally tested \cite{jensen_statistical_2018, korbel_classification_2018,korbel_thermodynamics_2021}. Additionally, as the monomers become more distinguishable the initial entropy of the system will increase reducing the total change. 

The times required for these experiments is usually on the orders of days \cite{ferus_formation_2017, mccollom_miller-urey_2013, parker_conducting_2014}. However, increasing the size of the experiment will increase the times required in a nonlinear way. Predicting the transient behavior of these systems is beyond the scope of this paper, as we have only determined the equilibrium solution through statistical means. The rate at which the aggregates form will increase like $p(n)$ due to the convergence of the limit shape for large $n$, but the number of reactions required for a single transition will grow rapidly as well on the order of $n^d$. Preliminary studies of time dependence are performed in studies by Kauffman et al. \cite{kauffman_theory_2020}. 

\section{Conclusion}
In conclusion, we view the process of the formation of aggregates from simple monomers through the lens of statistical mechanics. We take advantage of the existence of strong chaotic drives by making the ergodic assumption that the behavior of these systems will depend only on the multiplicity of states in the configuration space. We use the mathematics of integer and vector partitions to predict the limiting behavior of these systems for large times and large numbers of monomers. We provide analytic equations for predicting the maximum size aggregate and the entropy change involved. We find that the strength of aggregation increases non-extensively with the number of monomers and decreases with the distinguishability of monomers. These findings recast the process of the formation of complexity as a thermodynamic imperative. We then specifically apply these analytic results to the RNA world hypothesis. We show that our model justifies the formation of large molecules, such as strands of RNA, over geological timescales, something that current geometrically distributed models do not. We further give rough estimates of the required changes to current prebiotic experiments to produce more complexity. Specifically, increasing the scale of the experiment will allow more complexity to emerge. However, the ideal next step for this line of research is not to scale up the size of these experiments. First, the predictions of the limiting distributions of the partition models must be verified through experiments. In particular, the entropy of formation of these aggregates must be measured to determine whether or not it is extensive. This first experiment would require the reaction of monomers that are metastable, ideally with no energy incentive to bond and equal probabilities of bonding or breaking up. This experiment could be used to both measure the entropy generation and the limiting distribution. Additionally, modifications of the theory to include time dependence and the free energy of particular molecules must be included. The assumption of ergodicity must also be rigorously proven for these systems.

\section{Acknowledgments}
I would like to acknowledge Dr. Leonid Bogachev for useful discussions on the properties of the vector partitions. Additionally, I would like to thank Dr. Charles Rackson and Dr. Dileep V. Reddy for useful discussions.

\section{Disclaimer}
Certain equipment, instruments, software, or materials are identified in this paper in order to specify the experimental procedure adequately. Such identification is not intended to imply recommendation or endorsement of any product or service by NIST, nor is it intended to imply that the materials or equipment identified are necessarily the best available for the purpose.

\section{Appendices}
\subsection{\label{app:3} Numerical simulations of the vector partitions}

\begin{figure}
\includesvg[width=0.5\textwidth]{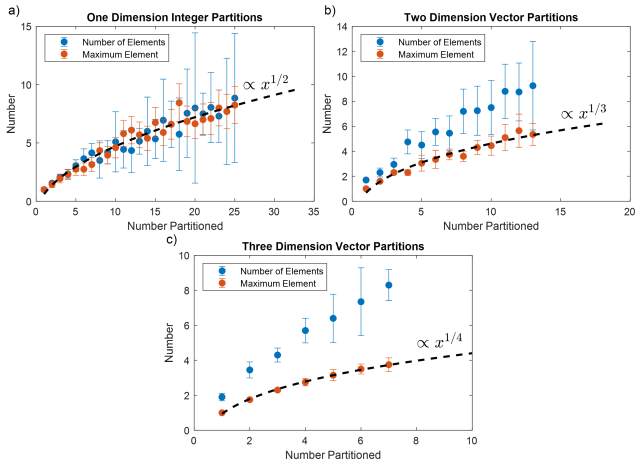}
\caption{\label{fig:sim} a) a depiction of the number of distinguishable elements from a partition of an integer N in blue and a depictition of the maximum element size in orange for 20 random partitions. A fit corresponding to $n^{1/(1+1)}$. b) a similar simulation for 2D vector partitions the slower increase of the maximum element is evident. c) a depiction of the same simulation for the 3D vector partitions. The random samples become difficult to calculate at the dimension with a conventional computer.}
\end{figure}

Simulations of the vector partitions were performed using SageMath 10.1.0 \cite{stein_sage_nodate}. The maximum element and number of elements for vector partitions of different sizes $n$ and dimensions $d$ were simulated using the random element function. Twenty random partitons of increasing sizes were sampled and averaged to produce the figure \ref{fig:sim}. Each figure corresponds to a different dimension of partition as labeled by the title. In the original paper outlining the vector partitions by Vershik the number of elements scaling is proposed to be $n^{1/(d+1)}$ while the maximum element scaling is given by $n^{d/(d+1)}$. The situation is actually the reverse where the maximum element scales slower than the number of elements. This makes sense as in the limit of $d=n$ the limit shape will become atomic, the bell measure.

\subsection{Limit shape derivation\label{app:1}}
This derivation was inspired by the variational method of maximizing random walks \cite{kenyon_limit_2007, cohn_variational_2000}. The limiting shape of the integer partitions corresponds to the most likely partition of a large integer. We can recast this problem into what is the limiting shape of a random Young diagram.
\begin{equation}
    \phi_{\lambda}(t) = \frac{\gamma_{n}}{n} \phi_{\lambda}(\gamma_{n}t)
    \label{eq:young_diag}
\end{equation}
such that $\int_{0}^{\infty} \phi_{\lambda}(t) dt =1 $. In general this scale factor is equal to $n^{1/(2+\beta)}$, where $\beta$ is the dimensionality of our integer partition $\beta = (d-2)/2$ for $ d\geq 2$ \cite{vershik_limit_2001}. We can now take the limit of large $N$ and obtain a continuous curve in the infinite limit (so long as the limit exists). This curve will correspond to the limiting shape of the partitions of an integer.

\begin{figure}
\includegraphics[width=0.5\textwidth]{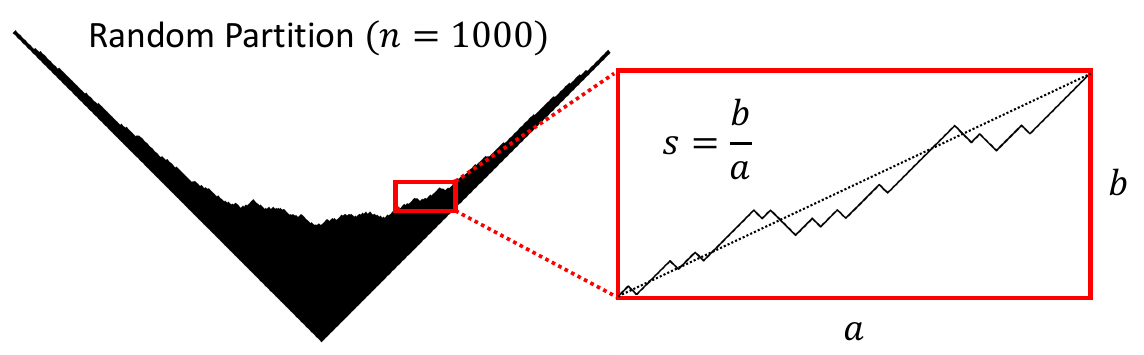} 
\caption{\label{fig:walk} A depiction of a random parition of 1000 rotated by 45 degrees. The red box depicts a zoom in on the surface of the partition. It is approximated as a random walk of length $a$ conditioned on two points}
\end{figure}

We will use a variational approach from \cite{okounkov_limit_nodate} to find a continuous function \(f(t)\) which in its neighborhood contains the most curves \(\phi_{\lambda} (t)\). Effectively a curve of maximum entropy. To do this, we will use the property that the curve \(\phi_{\lambda} (t)\) is monotonically decreasing, and that for small intervals corresponds to a random walk of \(a\) steps with a slope of \(s\). For simplicity of calculation, we will rotate the grid of the random walk by 45 degrees, but the proof is true in either case. Figure \ref{fig:walk} depicts a small section of the surface of a random partition as a random walk of length $a$ conditioned on two points generating a slope $s$. 

Many different random walks connect the lower right and top right corners of the box in Fig. \ref{fig:walk}. We can characterized them by where up steps are chosen. Out of the \(a\) total steps $u$ up steps must be chosen on any interval. Therefore the number of random walks that reach the same points in \(a\) steps with an average slope of \(s\) is

\begin{equation}
{a \choose u }
\label{eq:random_walk_binomial}
\end{equation}

We can calculate $u$ using the slope and total number of steps.  We know there are $b$ extra steps up, and then there are $(b-a)/2$ that get canceled out be down steps.
\begin{equation}
    b+\frac{a-b}{2}=u
\end{equation}
Using the slope $b=sa$ and simplifying we obtain 
\begin{equation}
{a \choose (sa+a)/2 }
\label{eq:random_walk_binomial}
\end{equation}
The total Using Stirling's approximation, we have that

\begin{equation}
{a \choose (sa + a)/2} \approx \exp(a H((s+1)/2))
\label{eq:sterling_approx}
\end{equation}

Where \(H(p)\) is just the entropy

\begin{equation}
H(s) = -p\ln(p) - (1-p)\ln(p)
\label{eq:entropy}
\end{equation}

Therefore, our function must maximize the functional

\begin{equation}
J[f] = \int \sigma\left(s\right) dt + c \int f(t) dt
\label{eq:functional}
\end{equation}

subject to the constraint of unit area, where \(\sigma\left(s(t)\right) = H((s+1)/2)\) and \(s=f'(t)\). Really \(f'(t)\) is the probability density function of the limit shape, so we are just maximizing the entropy of the integers distribution.

The Euler equations for this variational problem are then

\begin{equation}
c - \frac{d}{dt}\frac{\sigma(s)}{ds} = 0
\label{eq:euler_equation}
\end{equation}

Taking the derivative of \(\sigma\) we obtain

\begin{equation}
c - \frac{d}{dt}\left[\ln\left(\frac{1}{2}-s\right) - \ln\left(\frac{1}{2}+s\right)\right] = 0
\label{eq:sigma_derivative}
\end{equation}

Solving this differential equation we obtain

\begin{equation}
f(t) = \frac{t}{2} + a - \frac{\ln(1 + \e^{ct+b})}{c}
\label{eq:f_solution}
\end{equation}

or implicitly

\begin{equation}
y = \frac{x}{2} + a - \frac{\ln(1 + \e^{cx+b})}{c}
\label{eq:implicit_f_solution}
\end{equation}

The problem is unique up to scaling, so we can re-scale \(\frac{x}{2} \rightarrow x\). Solving for a simple implicit form we obtain

\begin{equation}
\e^{-cx} - \e^{a}\e^{-cy} = -\e^{b}
\label{eq:implicit_form}
\end{equation}

We require in the limit \(x \rightarrow \infty\) that \(y = 0\) and vice versa, so

\begin{equation}
0 - \e^{a}(1) = -\e^{b}
\label{eq:limit_condition1}
\end{equation}

\begin{equation}
1 = -\e^{b}
\label{eq:limit_condition2}
\end{equation}

The final constant is determined by integrating the limit shape and normalizing the area. The implicit equation for the limit shape is then

\begin{equation}
\e^{-\frac{x}{\pi\sqrt{6}}} + \e^{-\frac{y}{\pi\sqrt{6}}} = 1
\label{eq:limit_shape}
\end{equation}

This limit shape is the limit shape of the uniform partitions \cite{vershik_statistical_1996}. 

To calculate the limit shape for a particular $n$, both axes must be scaled by $\sqrt{n}$ \cite{vershik_limit_1994}. The limit shape of the partitions actually represents the cumulative distribution function of each group of parts. The probability distribution function is then $-\frac{dy}{dx}$ \cite{fatkullin_limit_2018}. This can be seen as the number of distinct parts of a certain size is $(y(i))-y(i+1)/((i+1)-(i))$, the length of the row subtracted by the length of the row above. 
\subsection{\label{app:2} 
Entropy of the partitions}

One can pick different quantities to define the microstate, such as the number of distinct summands, or the number of curves lying close to the limit shape. This will result in different distributions. In the case of the uniform measure on the partitions, the number of distinct parts follows a Gumbel distribution with variance proportional to \(\langle k^2 \rangle \propto n^{-1/2}\) and mean proportional to \(\langle k \rangle \propto \sqrt{n}\ln(n)\) \cite{vershik_fluctuation_2006}. The CDF of this distribution is

\begin{equation}
p(k) = p(n)\exp\left[-\frac{2 \e^{-\frac{k-\sqrt{n} \ln (n)}{2 \sqrt{n}}}}{c}\right]
\label{eq:gumbel_cdf}
\end{equation}

where \(c = \pi\sqrt{2/3}\). If we calculate the fraction of partitions lying between \(\langle k \rangle + \langle k^2 \rangle\) and \(\langle k \rangle - \langle k^2 \rangle\), we obtain a constant. While if we count the number of partitions around \(k=0\) or \(k=n\), we obtain \(0\) as \(n\) approaches infinity. Therefore, the entropy of a partition with \(\sqrt{n}\ln(n)\), or the limit shape, scales as \(\ln(p(n))\). If we include the multiplicity of the individual partitions according to the multinomial, for the typical partition we have

\begin{equation}
{\sqrt{n}\ln(n)} \choose r_{1},r_{2},...,r_{\sqrt{n}\ln(n)} = \Omega
\label{eq:multinomial}
\end{equation}

we obtain the multiplicity of the limit shape as

\begin{equation}
\Omega_{ls} = \Omega^{p(n)}
\label{eq:multiplicity_limit_shape}
\end{equation}

which will make the entropy scale as \(p(n)\ln(\Omega)\). The function \(\Omega\) is only dependent on \(n\) and the limiting distribution of the \(r_{k}\)s. Therefore, for most of the partitions, it will be constant. The multiplicity of the typical partition will be \(\Omega \approx 2^{n-1}/p(n)\). Therefore, the entropy will scale as \(p(n)(n-1)\ln(2)-\ln(p(n))\) or, to leading order, \(\exp{(\sqrt{n})}\).

The amount of free energy and force produced from producing the limit shape is given by \cite{roos_entropic_2014}

\begin{equation}
E = k_{B}T\Delta S = k_{B}T\Delta \left(\ln{p(n)}-S_{0}\right) = k_{B}T\Delta \left(\sqrt{n}-S_{0}\right)
\label{eq:free_energy}
\end{equation}

\begin{equation}
F = k_{B}T\frac{\partial S}{\partial n}
\label{eq:force}
\end{equation}

If the initial state is either a single aggregate of size \(n\) or \(n\) distributed particles, then the initial entropy is \(0\) and we have

\begin{equation}
E \propto \sqrt{n}
\label{eq:energy_single_aggregate}
\end{equation}

\begin{equation}
F \propto \frac{1}{\sqrt{n}}
\label{eq:force_single_aggregate}
\end{equation}

In the case of particles in which different configurations are distinguishable, then we have

\begin{equation}
E \propto \exp{\sqrt{n}}
\label{eq:energy_distinguishable_particles}
\end{equation}

\begin{equation}
F \propto \frac{\exp{\sqrt{n}}}{\sqrt{n}}
\label{eq:force_distinguishable_particles}
\end{equation}

In neither case is the energy extensive, In the first case the energy per particle decreases with increasing $n$. This means if the energy barrier to aggregation increases with $n$ then no aggregation will occur for large $n$. In the second case the energy per particle diverges for large $n$. This would imply that for large systems aggregation is inevitable, no matter what the barrier. It is currently unclear whether entropy can be non-extensive and means exists to convert such super-exponential entropy to an extensive quantity \cite{korbel_thermodynamics_2021, korbel_classification_2018}.


\bibliography{partitions} 

\end{document}